\documentstyle[aps,preprint,amssymb,psfig]{revtex}
\begin{document}
\title{Optical qubit by conditional interferometry}\author{Matteo G. A. Paris
\thanks{E-mail address:{\tt Matteo.Paris@pv.infn.it}}}\address{Optics Section, 
Blackett Laboratory, Imperial College, London SW7 2BZ, United Kingdom \\ 
Dipartimento 'A. Volta' and Unit\'a INFM, Universit\'a di Pavia, via Bassi 6, 
I-27100 Pavia, Italy} \date{\today} \maketitle
\begin{abstract} We suggest a method to prepare any chosen superposition 
$ a_0 | 0\rangle + a_1 |1\rangle $ of the vacuum and one-photon states.
The method is based on a conditional double-interferometer fed by an
one-photon state and a coherent state. The scheme involves only linear
optical elements and avalanche photodetectors, and therefore it should be
realizable with current technology. A realistic description of the
triggering photodetectors is employed, i.e. we assume that they can only
check, with a certain efficiency, whether or not any photon is present.  We
discuss two working regimes, and show that output states with fidelity
arbitrarily close to unit may be obtained, with non vanishing conditional
probability, also for low quantum efficiency at the photodetectors.
\end{abstract}
\section{Introduction}\label{s:intro}
The last two decades have witnessed a substantial development in quantum
engineering and measurement of light. Several kinds of nonclassical states 
of light can now be generated, and their quantum properties can 
be fully characterized by accessible measurement schemes \cite{spe,wel}. 
Besides fundamental interest, nonclassical states also find applications, 
as for example, the use of number states in quantum communication channels, 
and of squeezed light in high-precision gyroscopes and interferometers. More 
recently, the quantum engineering of light received new attention, which is 
mainly motivated by the potential improvement offered by quantum mechanics 
to the manipulation and the transmission of information \cite{ben}.  
Indeed, phenomena like teleportation \cite{tel}, and quantum dense 
coding \cite{qdc} found their first implementation in the quantum optical 
domain. \par
Photons do not interact, and this feature is very useful for 
the transmission of information without signal degradation. Indeed, the
typical figures for losses in optical fibers are below $0.3$ dB/Km.
On the other hand, the same characteristic poses limitations to the 
manipulation of the quantum information encoded into a quantum state 
of light. Photon-photon interactions needed for computation, in fact, 
take place only in active optical media, characterized by nonlinear 
susceptibility. Usually, such nonlinearities are small, or masked by the 
concurrent absorption processes. Only recently, new methods based on dark 
atomic resonance and electromagnetically induced transparency \cite{eit}
have been suggested to strongly enhance nonlinearity while suppressing 
absorption. The possibility of such giant nonlinearities renewed the interest 
for optical quantum technology, as it opens new perspectives to build 
single-photon quantum logic gates. \par 
In this paper, we devote our attention to the preparation of any chosen 
superposition of the vacuum and the one-photons states $a_0 |0\rangle 
+a_1 |1\rangle$. This is the simplest state of light that 
carries a complete phase information, and, in turn, it represents the 
simplest example of an optical qubit. Remarkably, this is a low-energy-expense 
encoding of quantum information, as it requires, in average, less than 
one photon for each qubit. In particular, for the conventional computational 
basis $|\psi_\pm\rangle=\frac1{\sqrt{2}}[|0\rangle\pm|1\rangle]$, i.e. for 
balanced superposition, half a photon for each qubit is required. \par 
Different methods have been discussed with the purpose of engineering 
superpositions of radiation states. Mostly, these are in the context 
of cavity QED, since the interaction with atoms passing through the cavity 
allows to select specific components of an initial signal \cite{epd}. 
More recently, an all-optical device, based a ring cavity coupled to the 
signal through a Kerr-medium, has been suggested to realize
Fock filtering, and thus preparation of superpositions \cite{fck}. 
In addition, a conditional scheme based on beam splitters and photocounters 
has been suggested to implement the optical state-truncation \cite{tru} of a
coherent state, which, in turn, is used to prepare  superpositions. As we will 
see, this last setup corresponds to a particular case of the present proposal. 
\par The present scheme involves only linear optical elements and avalanche
photodetectors, and therefore it should be realizable with current technology. 
In essence, it consists of two Mach-Zehnder interferometers arranged such
that one of the outputs from the first one is then used as input for the second
one. The first MZ is fed by an one-photon state, whereas the second is fed by
a weak coherent state. The output states of the second interferometers are 
then measured, and the conditional output state from the first MZ turns out 
to be a superposition of the vacuum and one-photon states. The amplitudes for
the two components can be tuned by varying the internal phase-shifts of
the two interferometers and the amplitude of the coherent input.\par
The paper is structured as follows. In the next Section the scheme is presented
and its dynamics is evaluated. Then, the ideal conditional output state is  
calculated for the case of a perfect photodetection process.  
In Section \ref{s:phr} we take into account the imperfections of realistic 
photodetectors, and study their effects on the preparation of the 
superposition. In the literature two models of photodetectors (PDs) have been
employed. In the first, which we use throughout the paper, 
it is assumed that PDs are only able to check, with a certain 
efficiency, whether or not photons are present. This is a reliable 
description of customary avalanche photodetectors, and we refer to this 
as {\sf YES/NO} photodetection. In the second model, PDs are still affected by 
non unit quantum efficiency, but now they are able to discriminate among the 
number of incoming photons, thus acting as photocounters. This description 
does not yet correspond to available PDs, as no evidence of detectors 
capable to discriminate between the presence of, say, $n$ and $n+1$ photons
for a generic $n$, have been reported.  
The photocounter model has been sometimes used in the literature 
on conditional measurements, remarkably in Ref. \cite{tru}, and this led 
to the conclusion that schemes are reasonably insensitive to the detectors' 
inefficiency. In general, this is no longer true when the realistic 
features of avalanche photodetectors are taken into account. 
On the other hand, we will show that the present scheme offers a working
regime in which the use of {\sf YES/NO} detectors is enough to assure the 
reliable preparation of any chosen superposition.  
Finally, Section \ref{s:out} closes the paper with some 
concluding remarks.
\section{Conditional double-interferometry}\label{s:cdi}
\begin{figure}
\psfig{file=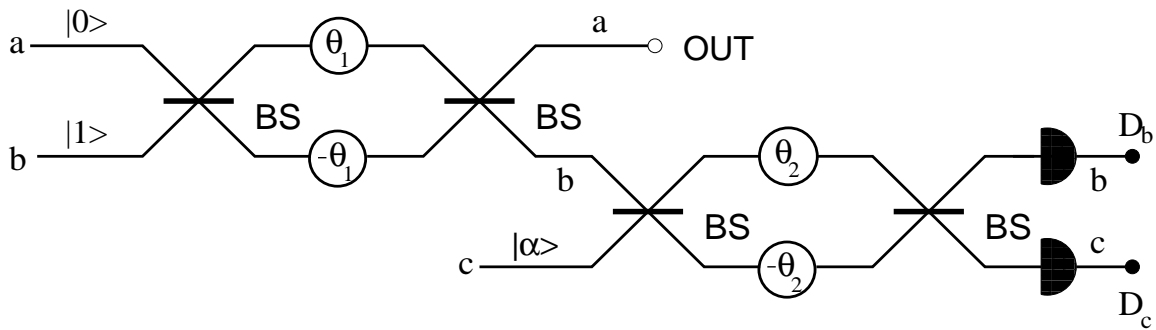,width=16cm}
\caption{ Schematic diagram of the conditional interferometric setup for the
preparation of any chosen superposition $ a_0 | 0\rangle + a_1 |1\rangle$ of
the vacuum and one-photon states. The BS' are identical balanced beam
splitters, whereas $D_a$ and $D_c$ denote two identical avalanche
photodetectors.  The first stage consists of a Mach-Zehnder interferometer
fed by an one-photon state in mode $b$. Then, one of the output from the first
interferometer is used as input of the other one, whose second port (mode $c$)
is excited in a weak coherent state. Both the output modes from the second
interferometer are detected by avalanche photodetectors, and depending on the
observed result we obtain different conditional output state in the mode $a$
(denoted by {\sf OUT} in the picture). The event of recording one photon in
one of the photodetectors (either $D_b$ or $D_c$) and no photons in the other
one corresponds to the preparation of a superposition of the vacuum and
one-photon states. The relative weights of the two components may be  tuned by
varying either the internal phase-shifts $\theta_1$ and $\theta_2$ or the
amplitude $|\gamma |$ of the coherent input, whereas the the phase of
superposition equals the argument $\arg\gamma$.
\label{f:set}}
\end{figure}

The scheme we have in mind  is the conditional double-interferometer depicted 
in Fig. \ref{f:set}. It consists of two Mach-Zehnder interferometers in 
cascade, in a way that makes one of the output signals from the first
one to constitute one of the input signals for the second one. The three 
field-mode involved in the setup are denoted by $a$, $b$, and $c$, whereas
the $BS$'s are symmetric beam splitters. We also assume that equal and 
opposite phase-shifts, denoted by $\theta_1$ and $\theta_2$ respectively, 
are imposed in the arms of each interferometer (see Fig. \ref{f:set}).
We assume that two interferometers are built with identical balanced beam 
splitters, and this means that they are fully characterized by the value of 
the internal phase-shift between their arms. The first interferometer is fed 
by an one-photon state in the mode $b$, whereas the other port, corresponding 
to mode $a$, is left unexcited. After this first stage the photon has a 
nonzero amplitude of being in both the output paths, whereas the values of 
such amplitudes are determined by the internal phase-shift $\theta_1$. The 
output mode $b$ is then mixed with mode $c$, prepared in a weak coherent 
state  $|\gamma\rangle$, and they are both detected at the output of the 
second interferometer. Depending on the result of the measurements we 
have different {\em conditional} output states for the output mode $a$ of 
the first interferometer. In particular, we will see that any chosen 
superposition $a_0 | 0\rangle + a_1 |1\rangle $ 
of the vacuum and one-photon states can be prepared with nonzero 
conditional probability.  \par
The evolution operator of each interferometer can be written as 
\begin{eqnarray}
\hat V_{\small MZ}(\theta )= \hat V_{\small BS}\: \exp\left\{i\theta
(a^\dag a-b^\dag b)\right\}\: \hat V^\dag_{\small BS} 
\label{mzevol1}\;,
\end{eqnarray}
where 
\begin{eqnarray}
\hat V_{\small BS} =\exp \left\{ i \frac{\pi}{4} (a^\dag b + b^\dag a)
\right\} \label{50bs}\;,
\end{eqnarray}
denotes the evolution operator of a symmetric beam splitter.
Eq. (\ref{mzevol1}) can be written as 
\begin{eqnarray}
\hat V_{\small MZ}(\phi)=\exp\left\{ i \frac{\pi}{2} b^\dag b\right\}
\exp\left\{-i\phi(a^\dag b+b^\dag a)\right\}
\exp\left\{-i\frac{\pi}{2}b^\dag b\right\}
\label{mzevol2}\;,
\end{eqnarray}
which shows that a Mach-Zehnder interferometer is equivalent to a single 
beam splitter $BS_\phi$ of transmissivity $\tau = \sin^2\phi$, where 
$\phi =\theta/2$, preceded and followed by rotations of $\pi/2$ performed 
on one of the two modes \cite{vis}. After straightforward algebra, one may 
write the evolution operator of the whole device as
\begin{eqnarray}
\hat U(\phi_1,\phi_2) =  \exp\left\{i\frac{\pi}{2}b^\dag b\right\}
\exp\left\{i\phi_2(b^\dag c + c^\dag b)\right\}
\exp\left\{i\phi_1(a^\dag b + b^\dag a)\right\}
\exp\left\{-i\frac{\pi}{2}b^\dag b\right\} \label{whole}\;.
\end{eqnarray}
The overall input state can be written as 
\begin{eqnarray}
|\Psi_{\sc in}\rangle =|0\rangle_a\:|1\rangle_b\:|\gamma\rangle_b
\label{psin}\;,
\end{eqnarray}
whereas using Eq. (\ref{whole}) we obtain the expression for the overall 
output 
\begin{eqnarray}
|\Psi_{\sc out}\rangle = \hat U (\phi_1,\phi_2) |\Psi_{\sc in}\rangle &=& \: 
\cos\phi_1 \: |1\rangle_a |\gamma\cos\phi_2 \rangle_b |\gamma\sin\phi_2
\rangle_c \nonumber \\ && +\sin\phi_1\sin\phi_2\: |0\rangle_a\:
b^\dag|\gamma\cos\phi_2\rangle_b |\gamma\sin\phi_2 \rangle_c 
\nonumber \\ && - \sin\phi_1\cos\phi_2 \: |0\rangle_a
|\gamma\cos\phi_2\rangle_b \; c^\dag|\gamma\sin\phi_2 \rangle_c 
\label{psiout}\;. 
\end{eqnarray}
We are now ready to analyze the effect on the mode $a$ of a measurement 
performed on the output modes $b$ and $c$. For the moment let us assume 
that the generic measurement of a quantity $X$ on the $b$ mode, and $Y$ 
on the $c$ mode is performed; we also assume that the two quantities are 
independent on each other. 
The measurement is thus described by a factorized probability operator
measure (POM) $\hat \Pi = \hat\Pi_x \otimes \hat\Pi_y$, where $x$ and $y$ 
denote the possible outcomes for the two quantities. The POMs are positive 
(hence selfadjoint) operators, if ${\Bbb X}$ and ${\Bbb Y}$ denote the spaces 
of the possible outcomes the normalization conditions can be written as  
($\widehat{\Bbb I}$ denotes the identity operator) $$\int_{\Bbb X}dx\:\hat\Pi_x
=\widehat{\Bbb I}\qquad\int_{\Bbb Y}dy\:\hat\Pi_y =\widehat{\Bbb I}$$.\par
The probability of the event $(x,y)$ is given by the global trace
over the three modes
\begin{eqnarray}
P_{xy}=\hbox{Tr}_{abc}
\left[|\Psi_{\sc out}\rangle\langle\Psi_{\sc out}|\; \hat\Pi_x\:
\otimes \hat\Pi_y\right]
\label{dprobxy}\;,
\end{eqnarray}
whereas the corresponding {\em conditional} output state for the mode
$a$ is given by the partial trace
\begin{eqnarray}
\hat\varrho_{xy} =\frac1{P_{xy}} \hbox{Tr}_{bc}
\left[|\Psi_{\sc out}\rangle\langle\Psi_{\sc out}|\; \hat\Pi_x\:
\otimes\hat\Pi_y\right]
\label{cond}\;.
\end{eqnarray}
Actually, different measurements on $b$ and $c$ lead to different
conditional output states, and since some events are more likely to 
occur than other, so are the corresponding conditional output states.
In the following we consider the measurement of the photon number, and 
in particular we focus our attention to the case of a single photon  
recorded in one of the output modes, and no photons recorded in 
the other one. In order to establish notation we assume that the
photon is recorded in the output mode $b$. However, the results are valid 
also for the case of of a photon registered in the mode $c$, up to a 
$\pi/2$ shift in the internal phase-shifts of the second interferometer.\par
The ideal measurement of the photon number on the two output modes is described 
by the POM 
\begin{eqnarray}
\hat\Pi_{nk} = |n\rangle_b\hbox{}_b\langle n|\otimes |k\rangle_c
\hbox{}_c\langle k| \qquad n,k=0,1,...
\label{ipomnk}\;.
\end{eqnarray}
Therefore, the detection probability for the case $n\equiv1$, $k\equiv0$ 
is equal to 
\begin{eqnarray}
P_{10} = \left|\langle\Psi_{\sc out}| 1\rangle_b|0\rangle_c\right|^2
= e^{-|\gamma|^2} \left[ \sin^2\phi_1\sin^2\phi_2 + 
|\gamma|^2 \cos^2\phi_1\cos^2\phi_2\right]\label{dprob10}\;,
\end{eqnarray}
and the conditional output state for the mode $a$ is given by
\begin{eqnarray}
\hat\varrho_{10} &=&\frac1{P_{10}} \hbox{Tr}_{bc} \Big[ 
|\Psi_{\sc out}\rangle\langle\Psi_{\sc out}|\; |1\rangle_b 
\hbox{}_b\langle1| \otimes |0\rangle_c \hbox{}_c\langle 0| \Big] =
\nonumber \\ &=&
\frac1P_{10}\left[
c_{00} \: |0\rangle\langle0| + c_{11} \: |1\rangle\langle1|
+ c_{01} \: |0\rangle\langle1| + c_{10} \: |1\rangle\langle0| \right] 
\label{cond10}\;,
\end{eqnarray}
where
\begin{eqnarray}
&& c_{00} = e^{-|\gamma|^2}\: \sin^2\phi_1\sin^2\phi_2 \qquad c_{11}=
e^{-|\gamma|^2}\:|\gamma|^2 \cos^2\phi_1\cos^2\phi_2\\
&& c_{01}= e^{-|\gamma|^2}\:\gamma\sin\phi_1\sin\phi_2\cos\phi_1\cos\phi_2
\qquad c_{10}=c_{01}^*
\label{ccoef}\;
\end{eqnarray}
the star denoting complex conjugation. By looking at 
Eqs. (\ref{ccoef}) it is easy to recognize that $\hat\varrho_{10}$ is
actually a pure state $\hat\varrho_{10}= |\psi_{10}\rangle\langle\psi_{10}|$
where
\begin{eqnarray}
|\psi_{10}\rangle= \frac{\sin\phi_1\sin\phi_2 |0\rangle + \gamma 
\cos\phi_1\cos\phi_2 |1\rangle}
{\sqrt{\sin^2\phi_1\sin^2\phi_2 + 
|\gamma|^2 \cos^2\phi_1\cos^2\phi_2}}
\label{psi10}\;.
\end{eqnarray}
By varying the interferometric shifts, and the amplitude $\gamma$ of the 
coherent input $c$, we may achieve any chosen superposition of the vacuum and 
the one-photon states. In particular, the internal phase-shifts $\phi_1$ and 
$\phi_2$, and the modulus $|\gamma|$, govern the weights
of the two component, whereas the relative phase equals the argument of 
the complex amplitude $\gamma$. \par We notice that the truncation scheme
of Ref. \cite{tru} is equivalent to a particular case of our setup,  
corresponding to the balanced choice $\phi_1=\phi_2=\pi/4$.
This is due, as mentioned above, to the fact that a Mach-Zehnder 
interferometer is substantially equivalent to a single beam splitter 
of transmissivity $\tau=\sin^2 \phi$. However, besides the fact that
the present scheme offers additional degrees of freedom, the use of an
interferometric setup has the specific advantage of a larger stability. \par
\begin{figure}
\psfig{file=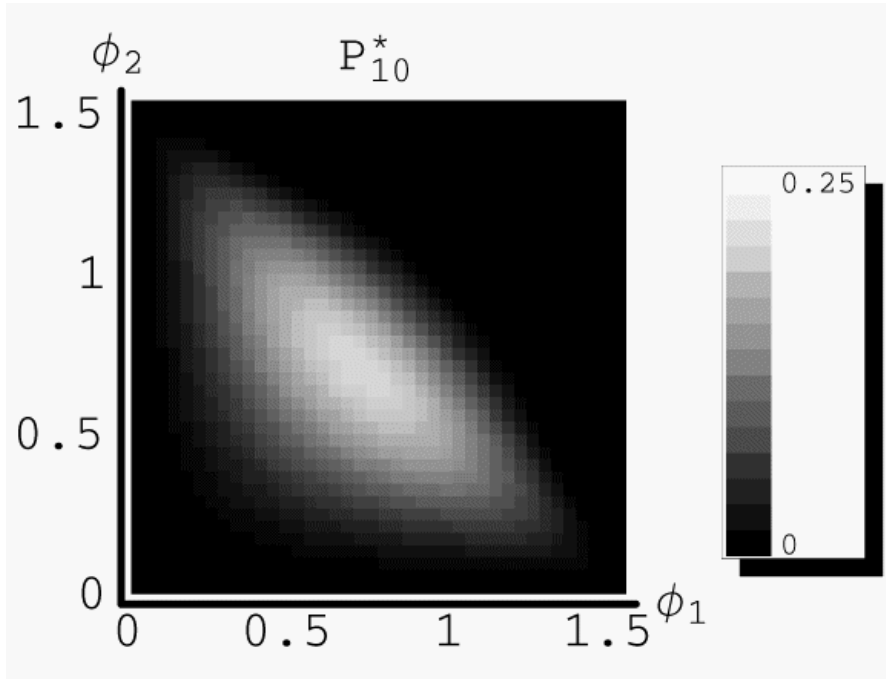,width=12cm}
\caption{Density plot of the conditional detection probability 
$P_{10}^\star$ for 
the preparation of the balanced superposition of Eq. (\ref{bal}). 
The maximum value ($P_{10}^\star\simeq 21\%$) is reached for 
$\phi_1\equiv\phi_2\simeq 0.715$, corresponding to an optimum amplitude 
$|\gamma|_{opt}\simeq 0.755$. However, the dependence of the detection 
probability on $\phi_1$ and $\phi_2$ is not dramatic, such, as it is apparent
from the plot, there exists a sizeable region in which the detection
probability is above $P_{10}^\star > 20 \%$. \label{f:pstar}}
\end{figure}

Remarkably, the superposition of Eq. (\ref{psi10}) may be obtained by 
different values of $\phi_1$, $\phi_2$ and $\gamma$, and this degree 
of freedom can be used to maximize the corresponding detection probability 
$P_{10}$, i.e. the probability of the event which leaves the state of the 
mode $a$ into the desired superposition.
Let us consider, for example, the preparation of a generic
balanced superposition 
\begin{eqnarray}
|\psi_\star\rangle = \frac{|0\rangle+e^{i\varphi}|1\rangle}{\sqrt{2}}
\label{bal}\;.
\end{eqnarray}
In order to reduce Eq. (\ref{psi10}) to Eq. (\ref{bal}) we need 
$\varphi=\arg \gamma$ and $|\gamma|=\tan\phi_1\tan\phi_2$. The detection
probability is then given by
\begin{eqnarray}
P_{10}^\star = 2\sin^2\phi_1\sin^2\phi_2\: \exp\big[-
\tan\phi_1^2\tan\phi_2^2\big]
\label{p10bal}\;.
\end{eqnarray}
From Eq. (\ref{p10bal}) we have that $0< P_{10}^\star \lesssim 0.21$, 
with the maximum value reached for $\phi_1\equiv\phi_2\simeq 0.715$, 
corresponding to an optimum amplitude $|\gamma|_{opt}\simeq 0.755$.
However, the dependence of the detection probability on $\phi_1$ and 
$\phi_2$ is not dramatic, such that there exists a sizeable region in 
which the detection probability is above $P_{10}^\star > 20 \%$ (see Fig.
\ref{f:pstar}). \par
As mentioned above, the symmetric case of a photon detected in the 
output mode $c$ and no photons in $b$ leads to an equivalent result, up to the 
replacement $\phi_2 \rightarrow \phi_2 + \pi/2$. In formula we have
\begin{eqnarray}
P_{01} = e^{-|\gamma|^2} \left[ \sin^2\phi_1\cos^2\phi_2 + 
|\gamma|^2 \cos^2\phi_1\sin^2\phi_2\right]\label{dprob01}\;,
\end{eqnarray}  and
\begin{eqnarray}
|\psi_{01}\rangle= \frac{\sin\phi_1\cos\phi_2 |0\rangle - \gamma 
\cos\phi_1\sin\phi_2 |1\rangle}
{\sqrt{\sin^2\phi_1\cos^2\phi_2 + 
|\gamma|^2 \cos^2\phi_1\sin^2\phi_2}}
\label{psi01}\;.
\end{eqnarray}
By comparing Eqs. (\ref{psi10}) and (\ref{psi01}) we also note that
the scalar product between the two conditional output state is given by
\begin{eqnarray}
\Big|\langle\psi_{01}|\psi_{10}\rangle\Big|^2 \propto
\sin\phi_2\cos\phi_2\:\Big(\sin^2\phi_1 - |\gamma|^2 \cos^2\phi_1\Big)
\label{over}\;,
\end{eqnarray}
which means that for $|\gamma| = \tan \phi_1$, and independently on 
$\phi_2$, the two states are orthogonal (this also happens for 
$\phi_2= p \pi/2$, $p\in {\Bbb Z}$ and any $\gamma$ and $\phi_1$, but this 
case just corresponds to have one state in the vacuum and the other in 
the one-photon state). In this regime, the scheme provides a 
reliable source (i.e. with $P_{10} = P_{01} = \exp\{-\tan\phi_1^2\} 
\sin^2\phi_1$ which is larger than $20 \%$ in the region around 
$\phi_1 \simeq 0.67$) of a quantum-optical computational basis.  \par 
Of course, it is now of interest to study whether or not the superpositions
of Eqs. (\ref{psi10}) and (\ref{psi01}) may be obtained in a realistic 
implementation of the setup. Since the scheme is based on conditional 
measurements, the main concern should be with the photodetection process. 
Therefore, in the next Section we are going to take into account the 
imperfections of available photodetectors, in order to check the robustness 
of the preparation scheme against the detectors inefficiency. 
\section{Effects of realistic photodetection}\label{s:phr}
Light is revealed by exploiting the interaction with atoms or molecules.  
Each photon ionizes a single atom, and the resulting charge is then amplified 
to produce a measurable pulse. In practice, however, available photodetectors 
are hardly performing the ideal measurement of the photon number. Their
performances, in fact, are limited by two main kinds of imperfections. On 
one hand, photodetectors are usually characterized by a quantum efficiency 
lower than unit, which means that only a fraction of the incoming photons 
lead to an electric pulse, and ultimately to a "count". Some photons are 
either reflected from the surface of the detector, or are absorbed without 
being transformed into electric pulses.  On the other hand, customary 
photodetectors involve an avalanche process to transform a single ionization 
event into a recordable pulse. This implies that it is very difficult to  
discriminate between the presence of a single photon or more than one. 
\par
The outcomes from such a detector may be either {\sf YES}, which means
a "click", corresponding to any number of photons, or {\sf NO}, which means
that no photons have been recorded. This kind measurement is described by
a two-value POM 
\begin{eqnarray}
\hat\Pi_{\sc N} =  \sum_{p=0}^{\infty} (1-\eta)^p \: |p \rangle\langle p|
\qquad  \hat\Pi_{\sc Y} = \widehat {\Bbb I} - \hat\Pi_{\sc N}
\label{yesno}\;, 
\end{eqnarray}
where $\eta$ is the quantum efficiency, and $\widehat {\Bbb I}$ denotes the 
identity operator. Indeed, for high quantum efficiency (close 
to unit) $\hat\Pi_{\sc N}$ approaches the projection operator onto the 
vacuum state, and $\hat\Pi_{\sc Y}$ onto the orthogonal subspace. \par
The event of observing a click at the PD surveying the output mode $b$ 
(i.e. $D_b$, see Fig. \ref{f:set}), and no photons at $D_c$, is 
characterized by the probability 
\begin{eqnarray}
P_{\sc yn}[\eta,\gamma,\phi_1,\phi_2] 
&=& \hbox{Tr}_{abc}\left[|\Psi_{\sc out}\rangle\langle
\Psi_{\sc out}|\; \hat\Pi_{\sc Y}\:\otimes \hat\Pi_{\sc N}\right] =
\nonumber \\ &=&e^{-\eta|\gamma|^2\sin^2\phi_2} 
\Bigg\{1- e^{-\eta|\gamma|^2\cos^2\phi_2} + 
\eta\sin^2\phi_1 \Big[e^{-\eta|\gamma|^2\cos^2\phi_2} \nonumber \\
&+& \cos^2\phi_2 \big(\eta |\gamma|^2\sin^2\phi_2 -1\big)
\Big]\Bigg\}\label{probyn}\;. 
\end{eqnarray}
The corresponding conditional output state is 
\begin{eqnarray}
\hat\varrho_{\sc yn} &=& \frac1{P_{\sc yn}} \hbox{Tr}_{bc}
\left[|\Psi_{\sc out}\rangle\langle\Psi_{\sc out}|\; \hat\Pi_{\sc Y}\:
\otimes\hat\Pi_{\sc N}\right] = 
\nonumber \\ &=& \frac1{P_{\sc yn}} \Big[ d_{00} \: |0\rangle\langle0| + 
d_{11} \: |1\rangle\langle1| + d_{01} \: |0\rangle\langle1| + d_{01}^* \: 
|1\rangle\langle0| \Big]
\label{rhoyn}\;,
\end{eqnarray}
where the coefficients are given by (see appendix A)
\begin{eqnarray}
d_{11} &=& e^{-\eta|\gamma|^2\sin^2\phi_2} 
\cos^2\phi_1\:\Big[1-e^{-\eta|\gamma|^2\cos^2\phi_2}\Big] \nonumber \\
d_{00} &=& e^{-\eta|\gamma|^2\sin^2\phi_2}
\sin^2\phi_1\: \Big[1- (1-\eta)e^{-\eta|\gamma|^2\cos^2\phi_2}
+\eta\cos^2\phi_2\big(\eta|\gamma|^2\sin^2\phi_2-1\big)\Big]\nonumber \\
d_{01} &=& e^{-\eta|\gamma|^2\sin^2\phi_2}
\: \eta \gamma \: \sin\phi_1\sin\phi_2 \cos\phi_1\cos\phi_2 
\label{coeffyn}\;.
\end{eqnarray}
In general, the conditional output state $\hat\varrho_{\sc yn}$ is no longer
a pure state. However, as we will see, there are regimes in which
$\hat\varrho_{\sc yn}$ approaches the desired superposition. In order 
to compare $\hat\varrho_{\sc yn}$ with the ideal conditional output 
$|\psi_{10}\rangle$ we consider the fidelity $F=\langle \psi_{10}|\hat
\varrho_{\sc yn}|\psi_{10}\rangle$. From Eqs. (\ref{psiout}) and 
(\ref{rhoyn}-\ref{coeffyn}) we have
\begin{eqnarray}
F[\eta,\gamma,\phi_1,\phi_2] &=& \frac1{P_{\sc yn}}\:\frac{e^{-\eta|\gamma|^2
\sin^2\phi_2}}{\sin^2\phi_1\sin^2\phi_2+|\gamma|^2\cos^2\phi_1\cos^2\phi_2}
\Bigg\{|\gamma|^2\cos^4\phi_1\sin^2\phi_2 \Big(1 - e^{-\eta|\gamma|^2\cos
\phi_2^2}\Big) \nonumber \\
&&+ 2\eta|\gamma|^2 \sin^2\phi_1\sin^2\phi_2\cos^2\phi_1\cos^2\phi_2+
\sin^4\phi_1\sin^2\phi_2\:\Bigg[1-(1-\eta)e^{-\eta|\gamma|^2\cos^2\phi_2}
\nonumber \\ &&+
\eta\cos^2\phi_2\big(\eta|\gamma|^2\sin^2\phi_2-1\big)\Bigg]\Bigg\}
\label{fid}\;.
\end{eqnarray}
Our goal is now to find regimes in which the fidelity of the conditional 
output state is close to unit and, at the same time, the corresponding 
detection probability  $P_{\sc yn}$ does not vanish. In particular, we 
are interested in the preparation of those superpositions where the 
amplitudes of the two components are of the same order, thus assuring 
that the state is far from being just the vacuum or the one-photon state. 
This requirement roughly corresponds to the condition
\begin{eqnarray}
\sin\phi_1\sin\phi_2\simeq\cos\phi_1\cos\phi_2
\label{balanced}\;,
\end{eqnarray}
whereas a  fine tuning of the amplitude, as well as the phase of the
superposition, may be obtained by varying the complex amplitude $\gamma$ 
of the coherent input. The condition in Eq. (\ref{balanced}) is satisfied 
by two different working regimes of the setup, i. e. by two different 
pairs of values of the internal phase-shifts. These are the balanced choice 
$\phi_1 = \phi_2 = \pi/4$ and the unbalanced one $\phi_1 \simeq 0$ $\phi_2 
\simeq \pi/2 - \phi_1$ respectively. In both cases, the general expression 
for the fidelity in Eq. (\ref{fid}) may be considerably simplified, and the 
corresponding working regime discussed with some details. 
\subsection{The case $\phi_1=\phi_2=\pi/4$}
\begin{figure}
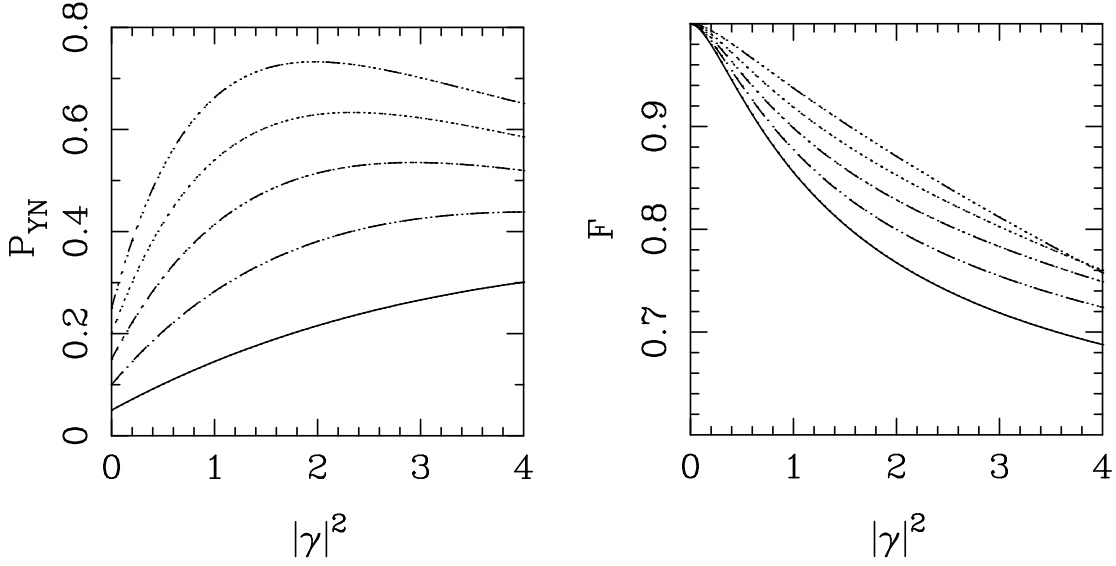

\begin{tabular}{ccc}
\psfig{file=fig3a.ps,width=7cm}&$\quad$&\psfig{file=fig3b.ps,width=7cm}
\end{tabular}
\caption{Performances of the setup with realistic description of the 
photo\-detectors: the case of balanced choice for the inter\-nal 
pha\-se-shifts. The figure shows the detection 
probability $P_{10}$ (on the left) and fidelity $F$ to the desired 
superposition (on the right) as a function of the intensity of the
input coherent state for different values of the quantum efficiency of
the photodetectors. In both plots we have, from bottom to top, $\eta= 20 \%
\hbox{(solid line)}, 40\%,60\%,80\%$, and $100\%$. 
\label{f:pynbal}}
\end{figure}

In the case $\phi_1 = \phi_2 = \pi/4$, the detection probability rewrites as
\begin{eqnarray}
P_{\sc yn}[\eta,\gamma,\pi/4,\pi/4] = e^{-\frac12\eta|\gamma|^2}\Bigg\{1-
e^{-\frac12\eta|\gamma|^2}+\frac12\eta\Big[e^{\frac12\eta|\gamma|^2}+\frac14 
\Big(\eta |\gamma|^2-2\Big)\Big]\Bigg\}\label{balpyn}\;,
\end{eqnarray}
and the fidelity
\begin{eqnarray}
F[\eta,\gamma,\pi/4,\pi/4] = \frac{\Big[4-2\eta+|\gamma|^2 (2+\eta)^2
\Big]- 4e^{-\frac12\eta|\gamma|^2} \Big(1-\eta+|\gamma|^2\Big)}
{(1+|\gamma|^2)\left[ 8+\eta (\eta 
|\gamma|^2-2) - 4 e^{-\eta|\gamma|^2/2} (2-\eta) \right]}\label{balfid}\;.
\end{eqnarray}
The detection probability shows relatively large values ($P_{\sc yn} 
\gtrsim 50 \%$) in a sizeable region (see Fig. \ref{f:pynbal}) 
of the $\eta-|\gamma|$  space, including also situations with low 
quantum efficiency. Unfortunately, the fidelity of Eq. (\ref{balfid}) 
is a rapidly decreasing function of the coherent amplitude and, 
in the relevant region $0<|\gamma|^2\lesssim 4$, it is bounded by 
\begin{eqnarray}
F_{\sc max} < 
\frac {2+|\gamma|^2 (9-4 e^{-|\gamma|^2/2})}{(1+|\gamma|^2)(|\gamma|^2 +6
-4 e^{-|\gamma|^2/2})}
\label{fidbound}\;.
\end{eqnarray}
This working regime is thus effective only for $|\gamma| \ll 1$, 
corresponding to the preparation of superposition where the vacuum
component is preponderant. Indeed, for balanced superpositions 
($|\gamma|\simeq 1$) we have the bound $F\lesssim 93 \%$. As we will see 
in the following, this limitation can be overcome by the unbalanced tuning 
of the  internal phase-shifts. \par 
Our analysis of the balanced scheme led to conclusions that are in contrast 
with those of Ref. \cite{tru}, where, as mentioned in Section \ref{s:cdi}, 
a formally equivalent scheme has been used. The reason for this disagreement 
stays in the different models employed to describe the photodetection process 
(see Appendix \ref{a:due}). Actually, as far as we know, the {\sf YES/NO} 
model used here is more realistic that the photocounter model used there, 
as, in fact, no evidence of detectors capable to discriminate between the 
number of incoming photons have been reported. We should conclude that the 
authors' hope of "reasonable insensitivity" to the detectors inefficiency 
\cite{tru} is not yet realized with current technology. \par 
\subsection{The case $\phi_1\simeq 0\:$, $\phi_2\simeq\pi/2-\phi_1$}
In the case $\phi_1 \simeq 0$, $\phi_2 \simeq \pi/2 
- \phi_1 $, the  detection probability is given by 
\begin{eqnarray}
P_{\sc yn}[\eta,\gamma,\phi_1\simeq 0,\phi_2\simeq\pi/2 -\phi_1] &\simeq &
e^{-\eta|\gamma|^2} \Big[\eta |\gamma|^2 \cos^2\phi_2 + \eta \sin^2\phi_1\Big]
\nonumber \\ &\simeq &  \eta\: \phi_1^2\: (1+|\gamma|^2) \: \exp 
\bigg[-\eta|\gamma|^2\bigg]
\label{unpyn}\;,
\end{eqnarray}
and the fidelity
\begin{eqnarray}
F[\eta,\gamma,\phi_1\simeq 0,\phi_2\simeq\pi/2-\phi_1] \simeq
1 - \frac12 \phi_1^2 (2-\eta)\label{unfid}\;.
\end{eqnarray}
Remarkably, the fidelity is now independent on the amplitude of the 
coherent input, and can be made arbitrarily close to unit by choosing
a smaller value for $\phi_1$. The price to pay for this result is a
lower value of the detection probability, namely a lower efficiency
of the preparation scheme. However, the resulting probability is still
large enough to make the scheme an effective source of superposition states.
As an example, let us consider $\phi_1$ such
that $\sin^2\phi_1=\cos^2\phi_2\simeq 0.01$. In this case, we obtain a very
high value of the fidelity $F >  99 \%$, and yet a detection 
probability given by $P_{\sc yn} \simeq 1\%$ (almost independently on the 
quantum efficiency). More generally, we can substitute Eq. (\ref{unpyn}) 
in Eq. (\ref{unfid}) to write
\begin{eqnarray}
F \simeq 1 - \frac{2-\eta}{2\eta(1+|\gamma|^2)} e^{\eta|\gamma|^2}
P_{\sc yn } \label{unrel}\;.
\end{eqnarray}
Eqs. (\ref{unpyn}), (\ref{unfid}) and (\ref{unrel}) 
assure that a reliable generation 
of the desired superposition is achievable (with non vanishing probability)
also for low quantum efficiency at photodetectors. \par
\subsection{Balanced superpositions}
\begin{figure}
\begin{tabular}{ccc}
\psfig{file=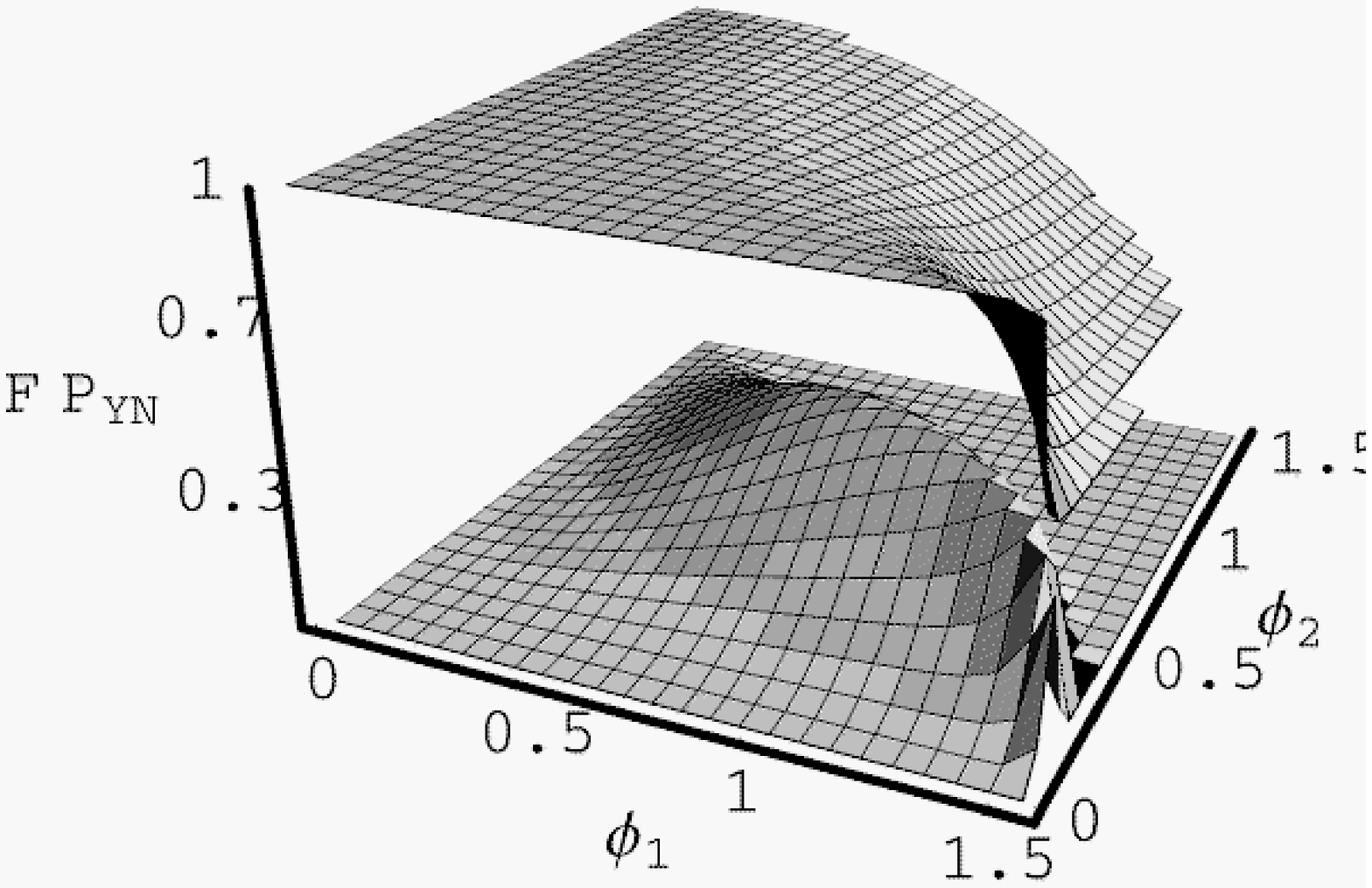,width=8cm}&$\quad$&\psfig{file=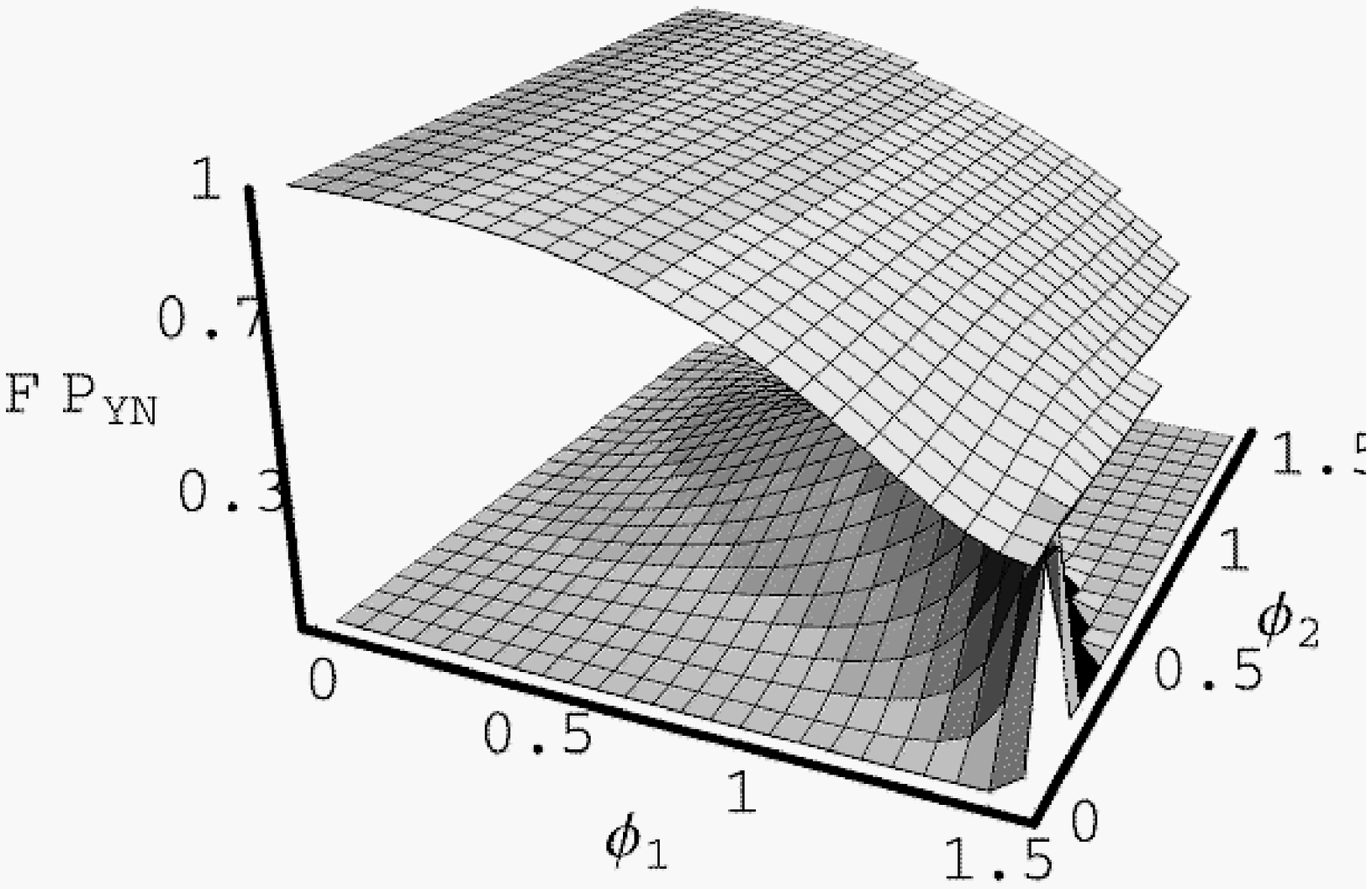,width=8cm}
\end{tabular}
\caption{Performances of the setup with realistic description of the 
photo\-detectors: preparation of balanced superpositions. 
The figure shows the detection 
probability $P_{10}$ and the fidelity $F$ to the desired 
superposition as a function of the internal phase-shifts 
$\phi_1$ and $\phi_2$ for unit quantum efficiency (on the left) and 
for $\eta=50\%$ (on the right).\label{f:fbal}}
\end{figure}

We end the Section by illustrating the performance of the setup in preparing 
the special class of {\em exactly} balanced superpositions of Eq. (\ref{bal}). 
The requirement for equal amplitudes reads $|\gamma|=\tan\phi_1\tan\phi_2$. 
By substitution in Eqs. (\ref{probyn}) and (\ref{rhoyn}) we obtain the 
detection probability and the corresponding conditional output state. 
We do not show here the resulting expressions, which are rather cumbersome.
Instead, in Fig. \ref{f:fbal}, we report the behavior of both, the fidelity
and the detection probability, as a function of the internal phase-shifts for
two values of the quantum efficiency at the photodetectors. As it is apparent
from the plots, there always exists a region, in which the fidelity is very 
close to unit, and yet the detection probability is larger than $10\%$. 
Therefore, for balanced superpositions, the performances with realistic 
detectors do not substantially differ from that obtained in the ideal working 
regime discussed in Section \ref{s:cdi}.
\section{Conclusions} \label{s:out}
We have analyzed a linear, conditional, interferometric setup to prepare any
chosen superposition $a_0 | 0\rangle + a_1 |1\rangle$ of the vacuum and
one-photon states. It consists of a three-port double-interferometer
fed by a one-photon state and a coherent state. The scheme involves only linear
optical elements and avalanche photodetectors, and therefore should be of
interest from the point of view of the experimental realization. In principle,
i. e.  in case of a perfect photodetection process, the setup can be used to
generate any chosen superposition with a conditional probability about $20\%$.
The imperfections of realistic photodetectors have been taken into account,
and their effects have analyzed in details.  An optimal working regime has
been found, in which output states arbitrarily close to the desired
superposition is obtained with non vanishing conditional probability. Typical
values for the fidelity are above $F \geq 99\%$, with conditional
probability about $P\simeq 1\%$.  For the relevant case of balanced
superposition, the detection probability may be increased by a fine tuning of
the amplitude of the coherent input.  In this case, the performances of the
setup are approaching the ideal working regime.
\section*{Acknowledgments}
This work has been cosponsored by C.N.R. and N.A.T.O. through the Advanced 
Fellowship Program 1998. The author thanks Peter Knight and Martin Plenio 
for their kind hospitality at Imperial College.

\appendix 
\section{Conditional density matrix for realistic 
photodetection}\label{a:uno}
Starting from Eqs. (\ref{psiout}), (\ref{cond}) and (\ref{yesno}) we have
\begin{eqnarray}
d_{11} &=& \cos^2\phi_1\langle\beta|\hat\Pi_{\sc y}|\beta\rangle 
\langle\delta|\hat\Pi_{\sc n}|\delta\rangle \nonumber \\
d_{00} &=& \sin^2\phi_1 \Big[ \sin^2\phi_2
\langle\beta|b \hat\Pi_{\sc y} b^\dag |\beta\rangle
\langle\delta|\hat\Pi_{\sc n}|\delta\rangle 
+ \cos^2\phi_2\langle\beta|\hat\Pi_{\sc y}|\beta\rangle
\langle\delta|c\hat\Pi_{\sc n}c^\dag|\delta\rangle \nonumber \\ 
&-&  \sin\phi_2\cos\phi_2 \Big( \langle\beta|\hat\Pi_{\sc y}b^\dag|\beta
\rangle\langle\delta|c\hat\Pi_{\sc n}|\delta\rangle+\langle\beta|b 
\hat\Pi_{\sc y}|\beta\rangle\langle\delta|\hat\Pi_{\sc n}c^\dag|\delta
\rangle\Big)\Big]\nonumber \\ d_{01} &=& \sin\phi_1\cos\phi_1
\Big(\sin\phi_2 \langle\beta|\hat\Pi_{\sc y}b^\dag|\beta\rangle
\langle\delta|\hat\Pi_{\sc n}|\delta\rangle 
+ \cos\phi_2\langle\beta|\hat\Pi_{\sc y}|\beta\rangle
\langle\delta|\hat\Pi_{\sc n}c^\dag|\delta\rangle \Big)
\label{ac1}\;
\end{eqnarray}
where $\beta=\gamma\cos\phi_2$ and $\delta=\gamma\sin\phi_2$.
Let us denote by $|z\rangle$ a generic coherent state with complex
amplitude $z\in{\Bbb C}$, then by using the definition (\ref{yesno}) of 
the POM $\left\{\hat\Pi_{\sc N},\hat\Pi_{\sc Y}\right\}$  we have
\begin{eqnarray}
\langle z |\hat\Pi_{\sc n}|z \rangle  &=& 
\exp\left\{-\eta|z|^2\right\}
\nonumber \\
\langle z |a \hat\Pi_{\sc n}|z \rangle &=& z (1-\eta) 
\exp\left\{-\eta|z|^2\right\}\nonumber \\
\langle z |a \hat\Pi_{\sc n} a^\dag |z \rangle &=& (1-\eta) 
\bigg[1+|z|^2 (1-\eta)\bigg]\exp\left\{-\eta|z|^2\right\}
\label{ac2}\;,
\end{eqnarray}
and
\begin{eqnarray}
\langle z |\hat\Pi_{\sc y}|z \rangle &=& 1 - \langle z |\hat\Pi_{\sc n}|z 
\rangle  \nonumber \\ \langle z |a \hat\Pi_{\sc y}|z \rangle &=& z - 
\langle z |a \hat\Pi_{\sc n}|z  \rangle  \nonumber \\ 
\langle z |a \hat\Pi_{\sc y} a^\dag |z \rangle &=& 1+ |z|^2 - 
\langle z |a \hat\Pi_{\sc n} a^\dag |z \rangle 
\label{ac3}\;.
\end{eqnarray}
Eventually, upon inserting Eqs. (\ref{ac2}) and (\ref{ac3}) in Eq. (\ref{ac1}) 
we arrive at the expression (\ref{coeffyn}) for the coefficients of the 
conditional output state.
\section{Modeling detectors as photocounters}\label{a:due}
By modeling a detector as a photocounter we assume that it is able to
discriminate among pulses of different amplitudes, ideally corresponding 
to the different number of recorded photons. Actually, the number of "clicks" 
cannot be not the number of incoming photons, as the photocounter is 
characterized by a nonunit quantum efficiency $\eta$. The POM describing 
the measurement is given by a Bernoulli convolution of the ideal photon-number 
POM $\hat \Pi_n = |n\rangle\langle n|$. In formula, we have
\begin{eqnarray}
\hat \Pi _n^\eta = \sum_{k=n}^{\infty} \eta^n (1-\eta)^{k-n} 
\left(\begin{array}{c} k \\ n \end{array} \right) |k\rangle\langle k|
\label{etaN}\;.
\end{eqnarray}
Therefore, compared to the picture of detectors as avalanche 
photodetectors, we have that the operator probability for the vacuum 
detection is the same, i.e.  $\hat\Pi_{\sc n} = \hat\Pi^\eta_0$, whereas 
for the case of a single click
\begin{eqnarray}
\hat\Pi^\eta_ 1 = \frac{\eta}{1-\eta} \sum_{k=1}^{\infty} k\: (1-\eta)^k
\: |k\rangle\langle k|
\label{1pd}\;.
\end{eqnarray}
Eqs. (\ref{ac3}) are now transformed into  
\begin{eqnarray}
\langle z |\hat\Pi^\eta_1|z \rangle &=& \eta |z|^2 e^{-\eta |z|^2}
\nonumber \\ 
\langle z |a \hat\Pi^\eta_1|z \rangle &=& \eta |z|^2 e^{-\eta |z|^2}
\left[1+|z|^2(1-\eta)\right]\nonumber \\ 
\langle z |a \hat\Pi^\eta_1 a^\dag |z \rangle &=& 
\eta |z|^2 e^{-\eta |z|^2} \left[ 1+ 3|z|^2(1-\eta) + |z|^4(1-\eta)\right]
\label{bc3}\;,
\end{eqnarray}
which leads to a detection probability given by
\begin{eqnarray}
P_{10}^\eta = \eta \Big[
\sin^2\phi_1 \sin^2\phi_2 + |\gamma|^2 \cos^2\phi_2 (1-\eta\sin^2\phi_1)
\Big]
\label{inaiveprob}\;,
\end{eqnarray}
and to a conditional output state whose coefficients are expressed as
\begin{eqnarray}
d_{11} &=& \eta |\gamma|^2 \cos^2\phi_1\cos^2\phi_2
\nonumber \\ d_{00} &=& \eta \sin^2\phi_1 \Big[
\sin^2\phi_2 + |\gamma|^2 (1-\eta) \cos^2\phi_2\Big] \label{naivecoeff} \\
d_{01} &=&  \eta\sin\phi_1\sin\phi_2\cos\phi_1\cos\phi_2 \Big[
1 + 2 |\gamma|^2 (1-\eta) \cos^2\phi_2\Big]\nonumber\;.
\end{eqnarray}
For the balanced setting $\phi_1=\phi_2=\pi/4$ we have 
$P_{10}^\eta = \eta/4 [1+|\gamma|^2 (2-\eta)]$ and
\begin{eqnarray}
d_{11} &=& \frac{|\gamma|^2}{1+|\gamma|^2(2-\eta)} \label{naivecoeff2} \\ 
d_{00} = d_{01} &=& \frac{1+|\gamma|^2(1-\eta)}{1+|\gamma|^2(2-\eta)}
\nonumber\;.
\end{eqnarray}
Finally, the fidelity is given by
\begin{eqnarray}
F[\eta,\gamma]= 1- \frac{|\gamma|^4
(1-\eta)}{(1+|\gamma|^2)[1+|\gamma|^2(2-\eta)]}
\label{naivefid}\;,
\end{eqnarray}
which is the result reported in Ref. \cite{tru}.
The fidelity of Eq. (\ref{naivefid}) varies in the range $5/6 \leq F \leq 1$ 
as a function of the quantum efficiency at the photodetectors.
\newpage 

\begin{thebibliography}{99}
\bibitem{spe} See, for example, the special issues: J. Mod. Opt. {\bf 44} 
(1997), on "Quantum state preparation and measurement", and Acta Phys. Slov.  
{\bf 48} (1998) on "Quantum Optics and Quantum Information".
\bibitem{wel} D.-G Welsch, W. Vogel, T. Opatrny, Prog. Opt., {\bf 39} (1999).
\bibitem{ben} C. H. Bennett, Phys. Scr. {\bf T76}, 210 (1998).
\bibitem{tel} D. Boschi, S. Branca, F. De Martini, L. Hardy, S. Popescu, 
Phys. Rev. Lett. {\bf 80}, 1121 (1998); S. L. Braunstein, H. J. Kimble, 
Phys. Rev. Lett. {\bf 80}, 869 (1998).
\bibitem{qdc} C. H. Bennett, S. J. Wiesner, Phys. Rev. Lett. {\bf 69}, 2881
(1992); K. Mattle, H. Weinfurter, P. G. Kwiat, A. Zeilinger, Phys. Rev. Lett.
{\bf 76}, 1895 (1996).
\bibitem{eit} H. Schmidt and A. Imamoglu, Opt. Lett. {\bf 21}, 1936 (1996); 
L. V. Hau, S. E. Harris, Z. Dutton, and C. H. Behroozi, Nature {\bf 397}, 
594 (1999); S. Rebi\'c, S. M. Tan, A. S. Parkins, D. F. Walls, J. Opt. 
{\bf B1},490 (1999). 
\bibitem{epd} J. Krause, M. O. Scully, T. Walther, and H. Walther, Phys. Rev. 
A {\bf 39}, 1915 (1989); M. Kozierowski, and S. M. Chumakov, Phys. Rev. A 
{\bf 52}, 4194 (1995);P. Domokos, M. Brune, J. Raimond, S. Haroche, Eur. 
Phys. J. D {\bf 1}, 1 (1998).
\bibitem{fck} G. M. D'Ariano, L. Maccone, M. G. A. Paris, M. F. Sacchi, 
Acta Phys. Slov. {\bf 49}, 659 (1999); see also quant-ph 9906077.
\bibitem{tru} D. T. Pegg, L. S. Philips, S. M. Barnett, Phys. Rev. Lett. 
{\bf 81}, 1604 (1998).
\bibitem{vis} M. G. A. Paris, Phys. Rev. A {\bf 59}, 1615 (1999).
\end{thebibliography}
\end{document}